# Plasmonic Photochemistry as a Tool to Prepare Metallic Nanopores with Controlled Diameter

German Lanzavecchia[a,b,†], Joel Kuttruff[c,†], Andrea Doricchi[a,d], Ali Douaki[a], Kumaranchira Ramankutty Krishnadas[e], Isabel Garcia[e,f], Alba Viejo Rodríguez[g], Thomas Wågberg[h], Roman Krahne[a], Nicolò Maccaferri*[g,h], and Denis Garoli*[a]

a. Optoelectronics Group, Istituto Italiano di Tecnologia, via Morego 30, I-16163, Genova, Italy.
b. Dipartimento di Fisica, Univesità di Genova, Via Dodecaneso 33, 16146 Genova, Italy
c. Universität Konstanz, Universitätsstr. 10, 78464 Konstanz, Germany
d. Dipartimento di Chimica, Univesità di Genova, Via Dodecaneso 31, 16146 Genova, Italy
e. CIC biomaGUNE, Basque Research and Technology Alliance (BRTA), 20014 Donostia-San Sebastián, Spain.
f. CIBER de Bioingeniería, Biomateriales y Nanomedicina, Instituto de Salud Carlos III
g. Department of Physics and Materials Science, University of Luxembourg, 106A Avenue de la Faïencerie, L-1511 Luxembourg, Luxembourg
h. Department of Physics and Umeå Centre for Microbial Research, Umeå University, SE-90187 Umeå, Sweden

*Corresponding authors: nicolo.maccaferri@umu.se; denis.garoli@iit.it;

**Abstract:** We show that plasmonic solid-state nanopores with tunable hole diameter can be prepared via a photocatalytic effect resulting from the enhanced electromagnetic field inside a metallic ring prepared on top of a dielectric nanotube. Under white light illumination, the maximum field intensity in these nanorings induces a site selective metal nucleation and growth. We used this approach to prepare bare Au and bimetallic Au-Ag nanorings and demonstrate the reduction of the initial inner diameter of the nanopore down to 4 nanometers. This process can be applied over large arrays with good reproducibility and good control on the nanopore diameter. The tunability of the nanopore diameter can be used to enable optimized detection of single entities with different size, such as single nanoparticles or biomolecules. As proof-of-concept, we demonstrate the versatility of the platform to perform single object detection of dsDNA, and Au nanoparticles with a diameter of 15 nm and 30 nm. We support our experimental findings with numerical simulations that provide insights into the electromagnetic field intensity distribution, showing that a field intensity enhancement of up to $10^4$ can be achieved inside the nanopores. This strong field confinement inside the final nanopore can be used to perform enhanced optical measurements, and to generate local heating, thereby modifying the ionic conductance of the nanopore.

Nanopore technology is the core of third generation sequencing, and solid-state nanopores are now one of the main topics in single molecule sensing. In order to make solid-state nanopores a valid alternative to biological nanopores, different nanofabrication methods and materials have been developed.[1-5] Electrical measurements are the main approach for single-molecule detection and sequencing by means of nanopores, but readout schemes that rely on optical spectroscopy have been proposed and demonstrated.[6-10] In this context, a plasmonic nanopore represents a unique tool to reduce the shot noise from the optical signal of small objects and molecules via the local enhancement of the electromagnetic (EM) field of the incident light used to detect the signal of these small entities. This EM field can be confined inside the diameter of the nanopore, down to the nm$^3$ volume, and engineered by leveraging on the size, shape and material composition of the nanostructure.[7] In the past 10 years, we have witnessed a growing interest in developing different types of plasmonic nanopores for single nanoparticle or molecule detection and sequencing, and more recently to achieve single protein resolution.[11-24,24-27] In most of the cases, the fabrication of sub-10 nm pores required complex multi-step processes, for example comprising the TEM sculpturing of the membrane where the plasmonic nanostructures are prepared.[28] Intriguing alternative strategies for plasmonic nanopore fabrication have been reported: for instance optical control of the dielectric breakdown of the membrane has been demonstrated by several groups,[29-31] and more recently the use of laser-driven fabrication of nanometric pores received significant interest within the community.[32-34] Nowadays, these approaches enable to prepare solid-state nanopores on dielectric or semiconductor membranes with good reproducibility, mainly on a single nanopore level. However, the preparation of nanopore arrays (towards parallel single molecule detection) with controlled diameter or the fabrication of nanopores in alternative materials such as noble metals is still challenging.[7,35,36] Methods for metallic nanopore fabrication comprise pore shrinking by means of metal growth or evaporation. Although this procedure is scalable to a large number of pores, it suffers from two major drawbacks: (i) it increases the thickness of the pore significantly, thus reducing the spatial confinement of the EM field; (ii) it is applied to the whole substrate. Therefore, metal deposition processes that can be spatially controlled on the nanoscale (without the use of lithographic processes) are highly desirable in the nanofabrication of plasmonic nanopores and represent a key advancement for this technology. A possible way to achieve this is by plasmonic photochemistry,[37-39] which refers to a synergistic combination of plasmonics and chemistry on the nanoscale to control metallic nucleation and growth. A pioneering work by Ai et al.[39] showed that the growth of Ag nanoparticles (NPs) can be controlled by means of plasmonic modes in the regions of maximum field enhancement in nanohole arrays. Following this idea, several interesting works have been reported recently.[40-45] In particular, Zhou et al. reported a similar experiment, where light was used to excite localized surface plasmons and drive a site-selective growth of Ag nanoparticles on nanobowl arrays.[44] Noteworthy, the high spatial control obtained for Ag growth was comparable or even below the diffraction limit of light due to nm-scale plasmonic field localization. The structures proposed in literature have also been used to perform enhanced spectroscopies[40,41,43,45] but so far they were not suitable for single molecule sensing, where configurations comprising nanocavities (such as zero mode waveguides, nanopores or metallic nanogaps) are preferred because of their extremely high EM field confinement.[46] Inspired by these recent results, here we demonstrate a scheme to prepare plasmonic nanopores as single object or in arrays with arbitrary configurations. Our design comprises a metallic ring on top of a dielectric tube as illustrated in Figure 1. The EM-field enhancement inside the ring is used to trigger localized metal nucleation and growth, thus reducing the diameter of the hole. We demonstrate that the process can be applied to different metals (for instance Au and Ag), albeit with different efficiencies. In particular, for Ag growth, the photocatalytic metal deposition tends to be slow (in the order of 1 nanometre per minutes), enabling a controlled reduction of the nanopore hole diameter down to 4 nm. A fine control of the size of the prepared nanopores is particularly important if the same platform concept is applied for the detection of different single entities. As a proof of concept, we demonstrate that the nanopores (or the arrays of them) fabricated with our photocatalytic approach can be used to detect relatively large objects such as nanoparticles (with size between 15 up to tens of nanometres) or small molecules such as DNA or protein (with size in the order of 1 nm). Numerical electromagnetic simulations are used to estimate both the EM-field enhancement and confinement inside the final nanopore, while electrical measurements enable to demonstrate the detection of different single entities (Au nanoparticles and DNA). Finally, the effect of optical excitation on the nanopore's conductance confirmed the plasmonic functionality that can be used also to perform enhanced spectroscopies, for example Surface Enhanced Raman Scattering (SERS) detection of a double-stranded DNA.

**Results and Discussion**

Figure 1 illustrates the concept of preparing metallic nanopores with nanometer precision in an ordered array or in arbitrary arrangement, by exploiting plasmonic photochemistry together with the ability of a metallic ring to confine and enhance the electromagnetic field in its inner part to enable nucleation and growth of an additional metal layer.

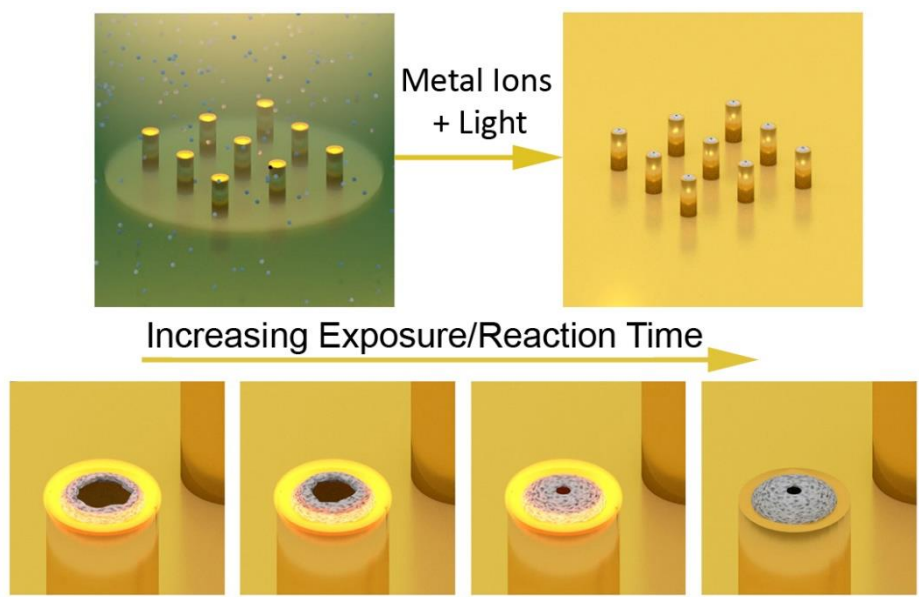

**Figure 1**. Illustration of the plasmonic photochemical reaction to prepare a metallic nanopore. An array of metallic rings is illuminated with white light and immersed in metallic salt solution.

Plasmonic effects can enhance the nucleation of metallic precursors[39] and allow to localize the seeding of nanoparticles in the regions of enhanced EM-field. We recently demonstrated the capability of a gold ring prepared on top of a dielectric tube to confine and enhance the EM-field.[47] Here we use this structure in combination with photochemistry for the nanopore fabrication. The structure consists of a hollow dielectric 3D tube with a 30 nm thick gold ring covering its top surface (see Methods for more details). Its dimensions, in terms of external and internal diameters depend on the fabrication procedures. Rings with external diameter between 200 and 600 nm can be prepared with high reproducibility. Correspondingly, the diameter of the axial hole can range from 60 to 300 nm. The EM-field is mainly confined in the inner part of the ring, with a maximum field intensity up to $10^3$.[47] A small EM-field enhancement is also present at the outer part of the Au ring with a maximum intensity up to $10^2$. Considering plasmon-driven metallic nucleation[39,41], it is reasonable to expect that such field confinement and enhancement can drive the local nucleation of metallic nanoparticles during a photochemical reaction. Hence, under illumination a significantly increased growth rate of metals can be expected due to the field enhancement, mostly in the inner part of the gold ring, and to a smaller extent on the external edges. In our experiments, the samples are illuminated with white light using an optical microscope, similar to the experiment reported by Ai et al. However, Ai et al.[39] applied the plasmonic photochemical growth to planar nanoholes on a glass substrate, which decorated the hole with Ag NPs, reaching minimal aperture diameters of 30 nm. In our work, we use a hollow nanopore configuration with a 3D geometry reaching aperture diameter of few nm. Moreover, we report a direct comparison between the 3D design and the planar configuration where the pores are prepared on a thin membrane.

**Nanopore Fabrication**

First, we verified the photochemical reaction in a planar configuration considering arrays of gold nanopores (with inner diameters of 70 nm) with different pitch between 400 and 850 nm. In this case, the EM-field in the inner part of the pores mainly depends on the pitch (periodicity) of the array as verified by finite element method (FEM) simulations (see Methods), using monochromatic light excitation (here @638 nm). The most effective pitch for EM-field confinement in the pore (see SI note#1 – Fig. 1S) is close to 600 nm. In order to experimentally verify the impact of the periodicity, arrays of nanopores with pitch varying between 400 and 800 nm were used as a platform to grow Ag via light excitation (both using monochromatic and white light sources – see Methods). We note that monochromatic light excitation is obviously more efficient in producing EM-field confinement in the nanoholes/nanorings, but white light source is easier to use in a standard optical microscope. As can be observed in Fig. S2 (SI – note#2), and in line with the FEM simulations, the most efficient plasmonic driven Ag growth is obtained for a pitch of 650 nm. However, the approach using this configuration did not result in reproducible Ag rings and did not provide the required control in size and uniformity over the array. This is partially in disagreement with the results obtained in Ref.[39], and is likely attributed to substrate effects, since in our case the Ag growth was performed on self-standing nanoholes. The absence of a substrate probably limits the efficiency of nucleation because of a smaller EM-field confinement and a different interaction between the metal salts and the substrate. However, the fabrication results obtained for the planar configuration are instructive and informative towards the design and expectations regarding the 3D nanotube-Au-ring.

We experimentally explored the fabrication of the proposed plasmonic nanopores on 3D nanostructures prepared in $Si_3N_4$ membranes. We first prepared the metallic rings on top of dielectric tubes. This fabrication relies on a well-established method firstly reported by De Angelis et al.[48] that was further developed during the last decade.[47,49,50] The versatility of the process allows to prepare 3D structures with high reproducibility, since the size (inner and outer diameters) depends only on the current used during the focused ion beam milling (see Methods). Figure 2(a,d) reports examples of gold rings prepared on top of dielectric tubes with different initial size (additional examples of array fabrication can be observed in SI note#3). The ionic current used for the focused ion beam fabrication ranges from 24 pA to 230 pA, and the obtained rings have external diameters between 200 nm to 550 nm, resulting in different resonant configurations for confined field in the inner part.[47] It is expected that the efficiency in metal ion nucleation during the plasmonic photochemical process depends on the dimension of the gold ring, since the process is related to the EM-field intensity. We compared the process using Au salts ($AuBr_3$ and $HAuCl_4$) and Ag salts ($AgNO_3$). In both cases, the structures have been illuminated with white light (Xe lamp, 50 W, focusing the light on the sample by means of a 20x (0.95NA) objective), and with a laser source (2 mW, @638 nm). During the exposure, the sample was immersed in the solutions for variable intervals of time between 10 and 50 minutes, in order to investigate the growth rate. The growth reaction was then stopped by transferring the sample into isopropyl alcohol and finally drying under $N_2$ flow. While for planar nanoholes the exposure with laser beam was possible, here it resulted in the destruction of the large part of the pillars (see SI – note#4). Therefore, we will report here the results obtained with white light excitation.

Both Au and Ag nucleation and growth lead to hole shrinking. Scanning electron microscopy (SEM) and Energy Dispersive Spectroscopy (EDS) have been performed to investigate the local deposition morphology and composition (Figure 2). EDS analysis confirmed that the Ag growth is localized on the pillar, with only negligible deposition on the metallic substrate (see also Fig. S8, SI – note#5). We note that for Au growth no contrast in EDS can be observed between Au on the substrate and Au growth on the pore.

To note, the hole shrinking rate observed for silver growth, using 5 mM $AgNO_3$ and Sodium Citrate as a reducing agent[39], was about 1 nm/min as shown Figure S9 (SI note#6)

Gold deposition was first performed via self-reduction of gold from $HAuCl_4$, which lead to a slow but rather inhomogeneous process resulting in a poor control of hole shrinking (SI – note#7). When replacing the $HAuCl_4$ with $AuBr_3$ salt and Sodium Citrate as reducing agent the reaction resulted to be fast with a full blocking of the pores after 15 min reaction time and with significant growth also on the substrate. Similar experiments were conducted with lower concentration of Au (1/10x and 1/100x) resulting in slower growth rate, but still significant metal deposition on the substrate was observed. At 0.5 mM Au precursor concentration, the pore shrinking rate was about 3 nm/min (SI – note#8).

The choice of a metallic nanoring on top of a dielectric pillar has an important impact on the spatial distribution of the metal growth. In principle, one could also use a fully Au covered 3D pillar for the growth, but then the EM-field confinement is extended along the full length of the hollow pillar (at resonance), and not limited to the top part of the pillar as for the metallic nanoring. [48,47,49,50] As expected, the seeding (due to plasmon-enhanced photochemical reactions) for fully Au covered pillars produced metal deposition all over the structure (see SI – note#9).

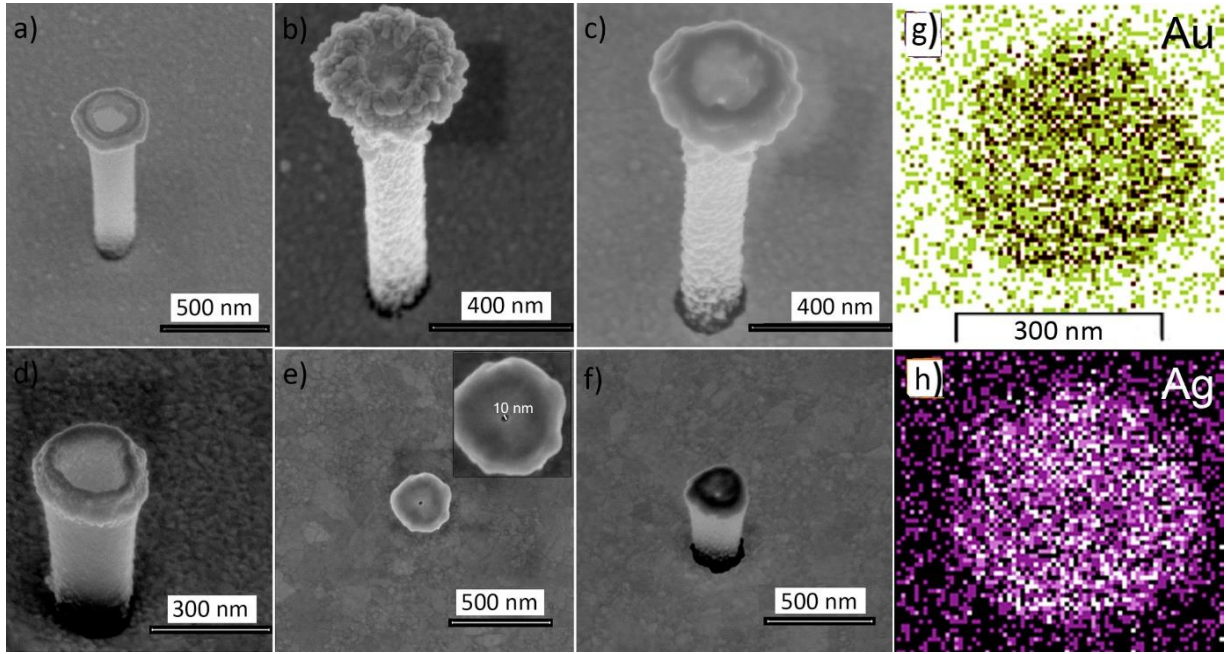

**Figure 2**. a-f) SEM micrographs of nanopores prepared via plasmonic photochemistry. a) and d) show examples of Au ring on top of dielectric pillars prepared at FIB currents of 40 and 80 pA, respectively; b) Morphology of the ring after 40 minutes of photochemical Ag reaction; c-f) Examples of the corresponding Ag nanopores (fig. a and d) morphology after thermal annealing; g-h) EDS maps of the ring for Au (g) and Ag (h) – The maps are referred to the sample illustrated in panel e-f.

The structures obtained and illustrated in Figure 2(b) (and in SI – note#3) demonstrate the localized growth of a metal obtained with optical excitation. The morphology appears very rough, and therefore with a relatively low control on the final shape and thickness of the grown layer. In order to obtain a smooth metallic structure, the samples have been annealed at 200°C for 10 minutes under $N_2$ flux (Figure 2(c,e,f) and Figure S14 – SI note#10). The annealing step is particularly effective in the case of Ag growth, where the large grains obtained after the photochemical reaction almost completely convert into a smooth metal film. The most important effect of the thermal annealing is that it enabled to obtain inner nanopore diameters smaller than 10 nm. Moreover, the thermal annealing also modified the shape of the metal ring producing a curvature in the inner part with a final bowl-like shape.

To verify that the metal deposition is indeed occurring by a photochemical process, we performed the reaction also without illumination. As illustrated in Fig. S15 after 30 minutes of reaction in the dark with $AgNO_3$, no Ag deposition was observed inside the Au rings (SI – note#11), hence demonstrating the key role of the plasmonic field confinement and enhancement.

**Numerical Simulations**

We performed finite element methods (FEM) simulations to obtain an estimation of the performance of our structures in terms of EM field confinement and enhancement. In detail, the RF Module in Comsol Multiphysics was used, taking the geometry of the hollow pillars coated with metal rings into account (Figure 3). The top view in Figure 3(a) shows that the EM field is mainly confined in the inner part of the ring, with a maximum field intensity up to $10^3$.

A small enhancement is also present at the outer part of the ring, where the maximum intensity reaches $10^2$.

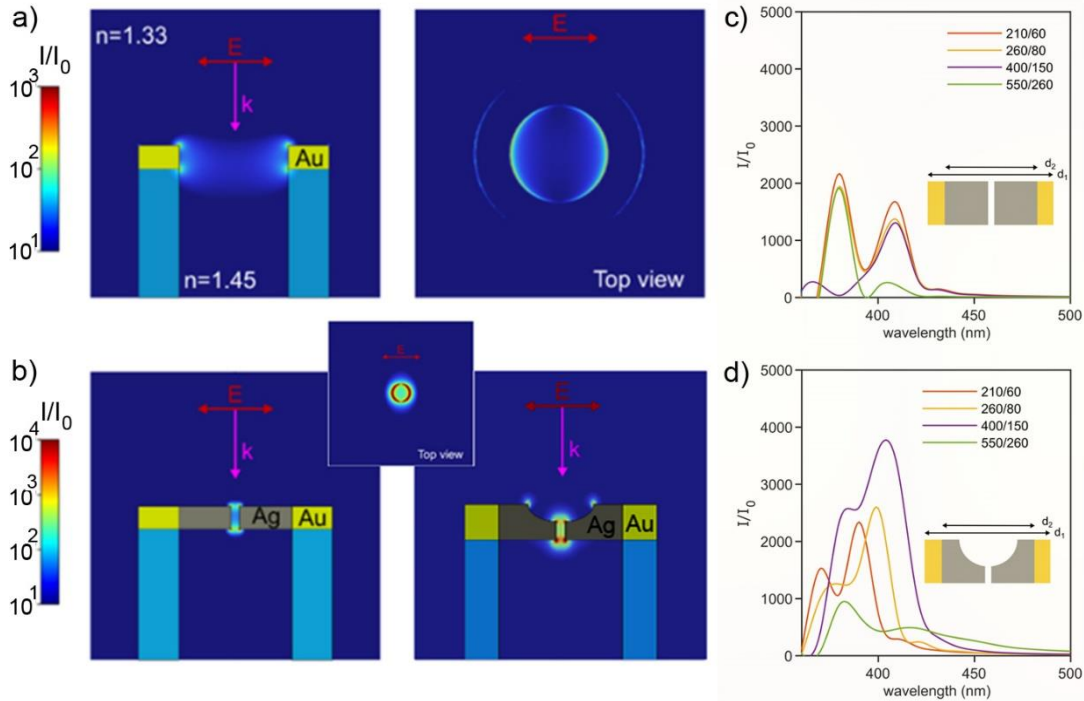

**Figure 3**. a) Electric field intensity enhancement $I/I_0$ with logarithmic color scale for the gold ring structure (outer diameter $d_1$=210 nm and inner diameter $d_2$=140 nm) at resonance condition ($\lambda$=665 nm). Incident k-vector and polarization are indicated by magenta and red arrows, respectively. b) The bimetallic Ag-Au ring with pore diameter of 10 nm at resonance ($\lambda = 410$ nm), left panel planar configuration, right panel bowl-like structure. c) Electromagnetic field intensity enhancement $I/I_0$ as function of the wavelength for the planar Au/Ag geometry using different sizes $d_1/d_2$ as indicated in the inset. The enhancement factor is evaluated as average on the top surface of the nanopore. d) same as (c) for the bowl-shaped geometry.

Figure 3(b-d) illustrates the behaviour of the structure after the inner diameter of the pore was reduced by the Ag deposition (simulations for Au deposition are reported in SI-note#12). Figure 3 shows that the reduction of the pore diameter leads to a stronger EM-field enhancement up to a maximum of $10^4$ at resonance (410 nm wavelength). Figure 3c shows the calculated spectral response $I/I_0$ of nanopores with an ideal geometry, and of the bowl-shaped metal rings (Fig. 3d) that are obtained in our fabrication (see Figure2(c)(e)(f)). The bowl-shaped structure obtained with the annealing display a stronger EM field enhancement (approximately twice of that of the planar structure illustrated).

Similar simulations for nanopores with additional Au deposition are reported in Fig. S16, S17 (SI - note #12)), which reveal a maximum field enhancement $I/I_0$ up to 150 at 610 nm excitation wavelength. The broad response of the Au nanopore in the visible spectral range suggests its potential use in enhanced spectroscopies. These latter can be also applied to Ag reduced nanopore although the functionality is optimized at lower wavelengths.

**Single Entity Detection and Electro-Optical Characterization**

To test the performance of our structures in different applications, in particular for single molecule detection, we investigated the nanopores with respect to the following aspects: i) the effect of illumination on the nanopore conductance; ii) the capability of the nanopore, prepared with different pore diameters, to detect single nanoparticles and biomolecules; iii) potential enhanced spectroscopy capabilities for parallel optical readout.

In order to characterize the electrical conductance of the nanopores, we fabricated single pores on a $Si_3N_4$ membrane, applying the photochemical reaction with Ag to reduce the hole of the nanopore to a diameter below 10 nm. The conductance was measured by I-V sweeping, and the data used to evaluate the hole diameter of the pore (Figure 4a; see SI – note#13 for details). The samples were placed in a microfluidic chip and immersed in TE buffer (10mM Tris-HCl containing 1mM EDTA•$Na_2$). Two AgCl electrodes have been used in combination with a nanopore reader[51] (see Methods for more details). In all cases the conductance of the nanopore has been measured before and after the photocatalytic metal deposition. We repeated these measurements on different samples and observed that a nanopore with a starting diameter of 60 nm resulted in a reduction by a factor of 10, down to 4 nm. Calculating this value it is important to consider the uncertainty in the final pore's thickness due to the fabrication procedure. In fact, while after the annealing step the pore thickness could be estimated about 30 nm, before that we can estimate a larger value (up to 150 nm). The minimum diameter of 4 nm correspond to a conductance of 4.2 nS (Figure 4a) and a pore thickness of 30 nm. The same conductance considering a pore thickness of 150 nm gives a pore diameter of 8.9 nm. It is important to stress that, although the preparation of a single nanopore with diameter of few nm has already been reported in literature [8,52–56], the controllable preparation of a metallic (plasmonic) nanopore array with diameter between tens of nm down to 5 nm is still very challenging. The method proposed here enables a controlled diameter reduction of the nanopore (1 nm/min with Ag growth) both in single pore configuration and in arrays with arbitrary pores. The EM field configuration in a solid-state nanopore has a fundamental role in the translocation process, and controlling the translocation process can enable significant improvements in molecule capturing, spectroscopic investigation,[7,36] and temporal resolution. Obtaining metallic nanopores with plasmonic properties that enable EM field engineering has the advantage that thermal effects can be explored. To investigate the thermal effects, a simple strategy could be to monitor the nanopore conductance with and without external light stimuli.[16,57,58] Thus, we measured the I-V response of the single plasmonic nanopore illuminated with LEDs emitting at different wavelengths (460 nm (blue) and 650 nm (red) – ca. 2mW power), respectively. In the Ag reduced nanopore, we expect to observe a more pronounced plasmonic effect with the blue LED. Our measurements, reported in Fig. 4(b), demonstrate an increased pore conductance for illumination with blue light (on-resonance) to G=10.9 nS. This is probably due to the EM-field enhancement produced by the Ag ring, which results in local plasmonic heating of the pore. Measurements off-resonance and without illumination lead to lower conductivity, G= 6 nS and 4.2 nS, respectively.

The fabrication of nanopores with controlled diameter is also interesting in optimized detection of the translocation events of single objects. In fact, the dynamic range of a nanopore measures the detectable size range of analytes. Since the sensitivity is closely related to the relative size of the analyte and the nanopore, the limited dynamic range of a single nanopore has remained an issue, which means it is difficult to resolve two molecules that differ by a few nanometers. Moreover, while biomolecules such as DNA or proteins requires nanopores with diameter of few nm, the detection of large NPs can be done only with larger nanopores. Here, we explored the electrical detection of dsDNA, and of Au nanoparticles (AuNPs) with size of 15 and 30 nm. This choice is aimed to the optimization of a platform that can be used in DNA nanostructures (comprising both DNA and nanomaterials) detection as in the case of DNA data storage as illustrated in our recent review and in refs.[59–62] The nanopore diameters used in two sets of measurements were evaluated by their conductance, yielding ca. 50 and 10 nm, respectively for AuNPs and DNA detection. The results of these experiments are illustrated in Fig. 4c and 4d. The DNA translocation data were first collected with a bandwidths of 200 kHz applying a constant voltage of 300mV (see Fig. S18 – SI note#14). Following the approach recently reported by Xia et al.[63] we used a Bessel filter (low pass) to a cutoff of 20 kHz for event analysis (Figure 4c). The signal collected at 200 kHz showed rather high noise (>1nA peak to peak) and this made the events detection more easy after a proper filtering. This high and low frequencies noise level can be due to several parameters, such as surface conditions or contamination[64,65] and the nanotube geometry. In particular, while the diameter of the nanopore is related to the final metal growth/annealing, the nanotube where the metallic pore is prepared is a dielectric channel on top of a $Si_3N_4$ membrane with a total thickness of hundreds of nm.[48] Comparing our results with recently reported DNA translocation in a $Si_3N_4$ nanopore it is possible to conclude that a higher noise is produced in our long channel tube+nanopore. After signal filtering, it was possible to clearly see the events of translocation and to calculate the histogram for the entire ionic current versus time trace (Fig. 4c). It shows two peaks: the baseline peak and a second peak corresponding to DNA events with a ⟨ΔI⟩DNA = 0.378±0.018 nA. This single peak related to translocation events suggests that our system was not able to detect folded and unfolded DNA configurations. This could be due to the large pore diameter used in the experiment (ca. 10 nm), but also to the not optimized signal to noise ratio.

In a second set of experiments, we used a nanopore prepared with a final diameter of ca. 50 nm to detect events of translocation of mixed AuNPs (with diameters of 15 and 30 nm). The NPs were prepared according to the protocol reported in methods and resulted as monodispersed NPs (see SI-note#15).They were first characterized in terms of ζ-potential and resulted to be slightly negatively charged (ca. -25mV). The translocation experiments were performed, with a bandwidth of 20 kHz, at different applied voltage observing zero or very few events of translocation for applied V<500mV. Translocation events, at a rate close to 0.01 event/msec, were detected with an applied voltage of 500 mV (Figure 4d). In this case, despite the low event rate, it was possible to observe two current levels (⟨ΔI⟩$AuNP_{15}$=0.22±0.03 and ⟨ΔI⟩$AuNP_{30}$=0.34±0.04) that can be associated to the two different NPs in the solution. The low event rate and the high voltage required to observe the translocation events can be ascribed to the low charge of the NPs and probably to surface charges configuration in the nanotube/nanopore. As described in the methods section, in order to avoid unpredictable surface charge configurations, the inner part of the nanotube/nanopore was coated with a 1.5 nm thick $Al_2O_3$ layer (with atomic layer deposition). This probably enabled to limit not uniform charges distribution between the metal layer and the dielectric substrate/nanotube, but geometrical aspects (due for example to the high aspect ratio of the tube) can play a role. For this reason alternative geometries should be considered, such for example a conical shape where it is reasonable to expect that the metallic nanopore can be prepared with the photocatalytic process with the same efficiency as reported here.

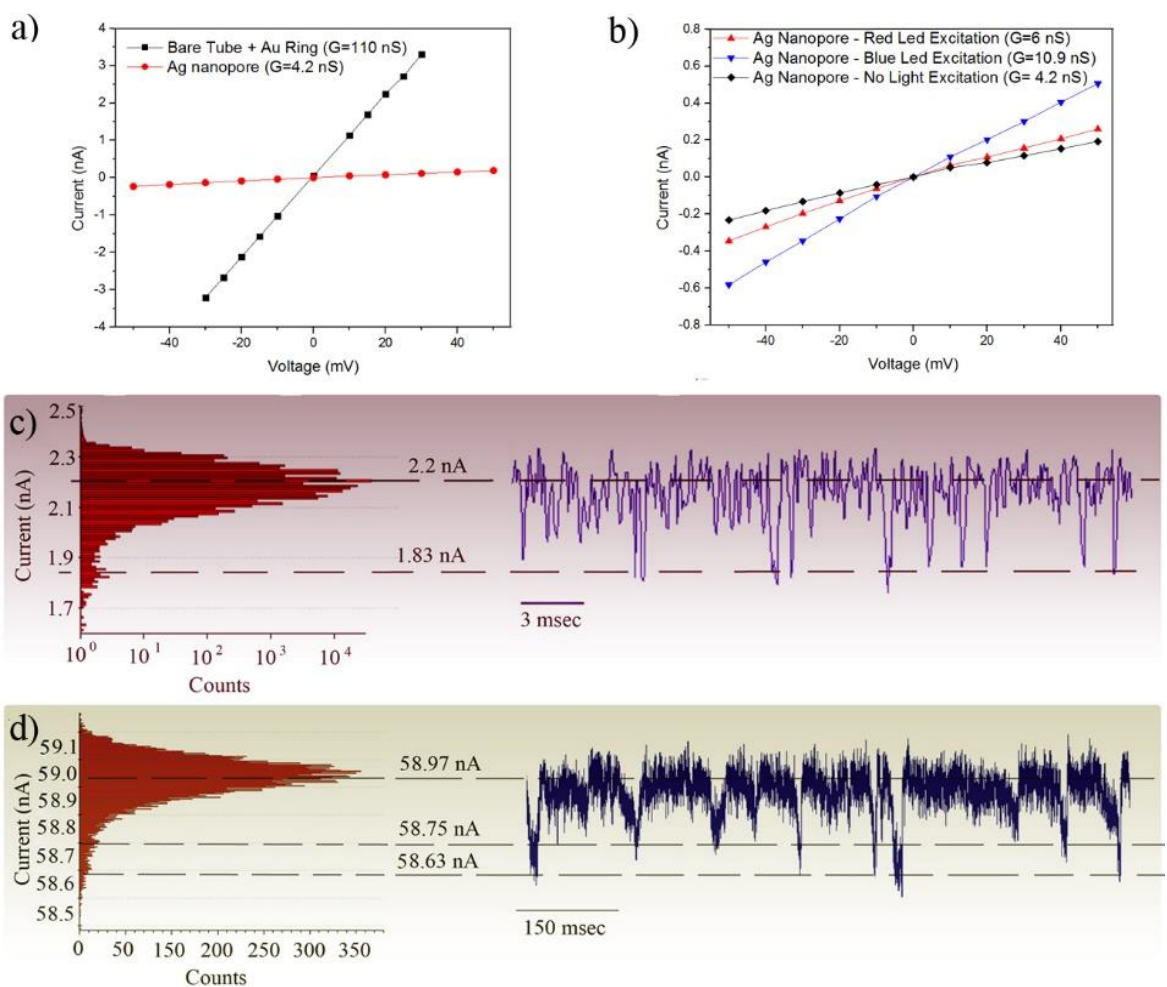

**Figure 4**. a) Current-voltage (I-V) curves obtained from a single nanopore with a Au ring, and the same nanopore after Ag deposition (40 minutes under white light illumination); b) I-V curves of a single Ag-Au nanopore illuminated with different light sources. c) dsDNA translocations in 1 M KCl at V = 300 mV. A 20 kHz low-pass Bessel filter is applied to distinguish event characteristics against noise. Translocation event shapes are consistent with single DNA translocation, although no clear discrimination between folded and unfolded states can be observed. d) AuNPs (15 and 30 nm) translocations in 1 M KCl at V = 500 mV were recorded at 20 kHz bandwidth using the portable nanopore reader from Elements SRL. Few translocation events can be detected per second but two distinct levels can be associated to 15 and 30 nm particles.

To evaluate our nanopore platform for field-enhanced spectroscopies, as a final proof-of-concept experiment, we applied the nanopore arrays for SERS by recording the signal of dsDNA (7nM) during the translocation through the pores. We used a setup similar to that reported in ref. [66], with laser excitation at 633 nm wavelength. For a direct comparison of the impact of the nanopore reduction, we compare the spectra obtained from nanopore arrays before and after the photocatalytic Au growth. We used Au deposition to be in resonance with the laser sources available in our Raman setup, since Au nanopores present a broad EM field enhancement between 550 and 750 nm (see Fig. S16, SI - note#12). Ag-Au nanopores are more efficient in the spectral range between 350 and 450 nm as shown in Figure 4. Representative Raman spectra are reported in Fig. 5, showing the two best signals obtained for the photochemically reduced pores and for the bare Au ring, respectively (details on the observed peaks are reported in SI-note#16).[67] With the Au nanopore (black curve Fig. 5a) it was possible to observe all the major peaks that can be associated to the dsDNA vibration, with higher efficiency with respect to the bare Au ring (red curve). The time map reported in Fig. 5a right panel, shows the low event rate for this experiment. In fact, over 500 seconds of data collection, only few tens of good spectra were recorded (with an integration time of 100 msec). This suggests a small dwell time as also observed in electrical recording. Anyway, the ability of the plasmonic platform to enhance the detection of the biomolecule translocation is confirmed. Finally, to corroborate the local field enhancement effect, we recorded spatial maps of the SERS intensity of Au nanopores array covered R6G as target analyte, which are reported in Figure 5. Raman spectra of R6G molecules (concentration 0.1 mM) acquired from the array of 3D nanopores substrate exhibit all the major lines at 611, 773, 1127, 1183, 1310, 1360, 1507, 1573, and 1648 cm$^{-1}$. Each of the Raman spectra was corrected by subtracting the fluorescence background and the clear different between spectra acquired from the nanopore area and the substrate demonstrates the enhancement due to the structure.[68]

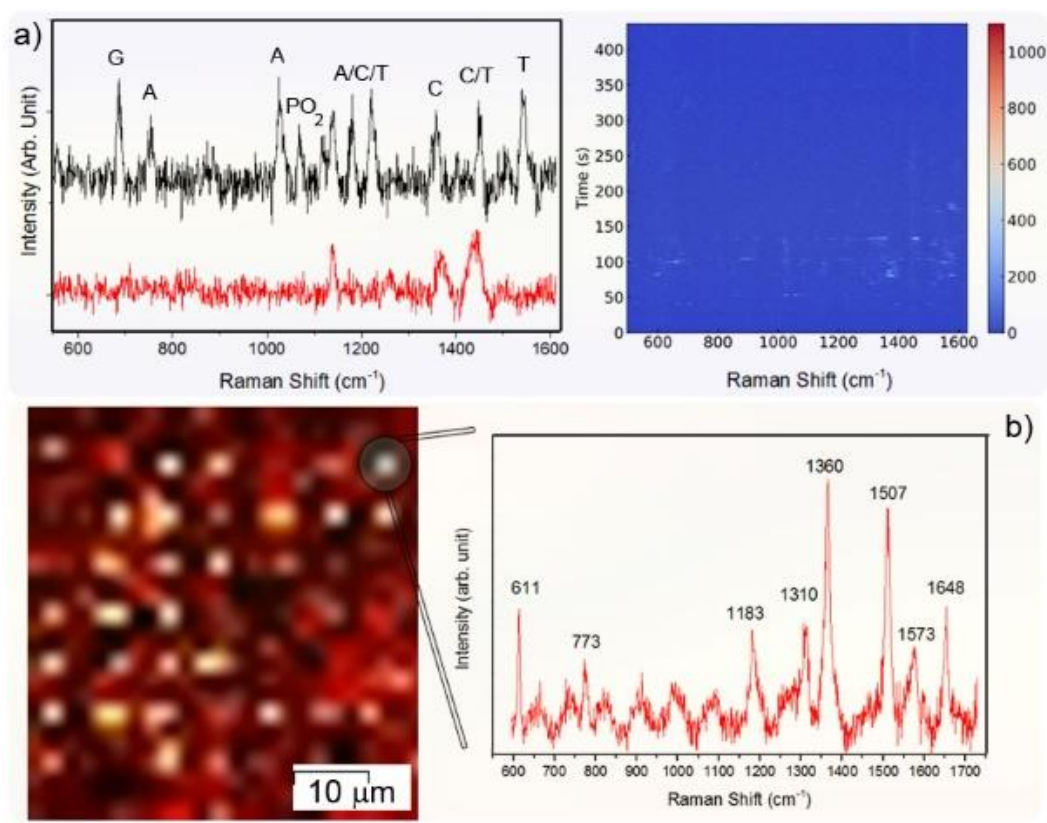

**Figure 5**. a) SERS spectra of dsDNA translocating through an Au ring on top of the pillar (black curve) and an Au nanopore (red curve) and time trace over 500 seconds. B) Map of the Raman spectra obtained from 3D nanopores array with R6G. The map is obtained integrating the peak at 1360 cm$^{-1}$. A detail on the full spectrum obtained from a nanopore is also reported.

**Conclusions**

In summary, we demonstrated that plasmonic photochemistry can be used for the controlled reduction of the diameter of plasmonic nanopores down to 4 nm. With respect to previously reported plasmonic nanostructures, our scheme significantly improves the control on the final nanopore's diameter. Another strength of our approach is that it can be applied to single nanopores, arrays of nanopores, or even arbitrary nanopore arrangements. The good control in nanopore size reduction over time (1nm/min with Ag) enables the fabrication of single or multiple nanopores with controlled diameter that can be optimized for the detection of objects with different size. Nanopores with sub-10 nm apertures lead to very strong EM field enhancement, which makes this platform appealing for field enhanced spectroscopies such as fluorescence, FRET and SERS[7].

Furthermore, we think that the plasmonic photochemistry approach that we presented is not limited to the metallic ring geometry or even nanopores, but that it can be applied to plasmonic structures in general that result in well-defined spatial EM field enhancement. Also, we believe that such an approach represents a valuable tool to investigate plasmon-driven photochemical reactions, charge transfer and other intriguing nanoscale chemical phenomena and an important step towards the realization of new plasmonic devices for several applications, from nanopore-based single-molecule detection and DNA or protein sequencing to ion selectivity and multiplexing information processing.

**Methods**

*Fabrication procedures*

Freestanding $Si_3N_4$ membrane chips were prepared following a standard membrane fabrication procedure. In particular, we prepared an array of square membranes a double sided 100 nm LPCVD $Si_3N_4$ coated 500 µm Si wafer via UV photolithography, following reactive ion etching and subsequent KOH wet etching.

Hollow nanopillars were fabricated by using a reported procedure[47]. In brief, properly diluted S1813 MICROPOSIT photo resist was spin-coated onto the $Si_3N_4$ chips. Then a thin layer (80 nm) of Au was deposited via sputtering. Focused Ion Beam patterning was used to expose the resist in arbitrary arrays by using different ion currents (24, 40, 80 and 230 pA, respectively). The choice for different ion currents allows for the preparation of different diameters in the final pillar/Au ring. After the exposure, the Au top layer was removed with gold etchant (KI solution) with a following oxygen plasma treatment (100 W 180s) to clean the upper part of the resist layer. A development in acetone with a further oxygen plasma (100 W 150s) produced the final hollow dielectric structure. A thin layer of Ti//Au (2//30 nm) was then deposited on top of nanopillars via electron beam evaporation.

Planar pores were fabricated with FIB on $Si_3N_4$ membranes previously coated with a thin layer of Ti//Au (2//50 nm).

The substrate was immersed in a Silver nitrate solution that was prepared with 30mL 10mM $AgNO_3$/ 1.8 mL 6mM Sodium citrate, and then illuminated with white light (Xe lamp 50 W / 20X 0.95NA objective) or 638 nm laser light for different durations[39]. For gold deposition, 5.0, 0.5 and 0.05 mM $AuBr_3$, and 10 mM $HAuCl_4$ solutions were used. Samples were then rinsed in isopropyl alcohol and dried under $N_2$ flow.

*Electrical characterization*

Electrical measurements for conductance estimation were conducted in 10 mM Tris-HCl/ 1 mM EDTA buffer solution, in the dark or under illumination with a LED at 650 nm, and 460 nm (power 2 mW) inside a nanopore reader using Ag/AgCl electrodes [51].

Electrical readout of DNA and NPs translocations were conducted in 1M KCl by using the same instrument / reader from Elements S.r.L. [63]

*Raman Spectroscopy*

Raman spectroscopy measurements were performed using a Renishaw InVia Raman system with a 100 × 0.95 NA water immersion objective. We collected for 500 seconds the signal from a dsDNA through the pore with an integration time of 0.1 s. In this case, an excitation wavelength of 632.8 nm were used for the experiment performed with Au ring and Au reduced nanopores, respectively. The dsDNA was 7 nM λ-DNA dispersed in 1M KCl. In order to enable the translocation, the solution was dropped in the bottom side of the membrane, while the upper side was filled with the buffer alone and moved close to the microscope objective used to excite the sample and collect the spectrum, a voltage of 100 mV was applied. A second experiment was performed by using R6G (0.1mM) dispersed in $H_2O$ without external applied voltage. The use of a resonant Raman molecule, excited at 632 nm, was chosen in order to rapidly test the in-array spectroscopy.

*Dynamic Light Scattering*

Dynamic light scattering experiments were performed by using a Malvern Zetasizer, and the measurements were evaluated by using Zetasizer software. Data are reported as the average of three measurements. The procedure used for these measurements followed the protocol reported in ref.[69]

*Numerical Simulations*

Numerical simulations were carried out using the finite elements method implemented in the RF solver of the commercial COMSOL Multiphysics software. The geometry was set up in 3D, and perfectly matched layers were introduced to suppress back reflections from domain walls. Linearly polarized light (plane wave) at different wavelengths was injected at the top of the simulation geometry under normal incidence to the structures. Interpolated data from Rakić et al.[70] was used to describe the linear optical properties of silver and gold. Refractive index of the dielectric pillars and the environment were set to 1.45 and 1.33, respectively

*Synthesis of 15 nm and 30 nm PEGylated Au NPs*

Chemicals.
Sodium borohydride (ReagentPlus, ≥99%, $NaBH_4$), poly(ethylene glycol) methyl ether thiol average Mn 2000 and 6000, L-ascorbic acid (ACS reagents, ≥99%, AA), hypochlorite (6−14% active chlorine), and cetyltrimethylammonium chloride (≥98%, CTAC) were purchased from Sigma Aldrich. $HAuCl_4·3 H_2O$ (≥99.9%, trace metal basis) was purchased from Alfa Aesar. All solutions, except $HAuCl_4$ and CTAB, were prepared immediately before use. Purified Milli-Q water was used in all experiments (Millipore, 18.2 MΩ cm). Glassware was cleaned with aqua regia and rinsed extensively with Milli-Q water before use.

Synthesis and Functionalization of Au NPs.
Gold nanospheres, Au15 and Au30, were synthesized according to the Turkevich (1) and seeded (2) methods respectively. For ligand exchange, the particles were mixed with thiolated PEG ligands in water (PEG2K for the case of Au15 and PEG6k for Au30 NPs), to obtain a final Au0 concentration of 1 mM (according to the absorbance at 400 nm) and a PEG-SH content of 1 mg/mL. After stirring overnight at room temperature, the excess of PEG-SH was removed by centrifugation at 12000 rpm (Au15) or 4800 rpm (Au30) and the pellets were redispersed in water.

*Nanoscale Characterization Techniques*
AuNPs were characterized by means of Transmission Electron Microscopy. TEM images were obtained using a JEOL microscope at an acceleration voltage of 200 kV. Approximately 3 μL of sample was dropped on a lacey carbon-coated grid and left to dry. The size distribution of the nanoparticles obtained was analyzed using ImageJ software

The nanopores were characterized by using a Scanning Electron Microscope (SEM) FEI Nova 600i equipped with EDS.

**Funding**

D.G. and R.K. acknowledge funding from the European Union under the H2020 Programme (No. FETOPEN-01-2018-2019-2020, Grant No. 964995 'DNAFAIRYLIGHTS'). N.M. acknowledges support from the European Union under the H2020 Programme (FETOPEN-01-2018-2019-2020 Grant No. 964363 'ProID'), the European Union under the Horizon Europe Programme (HORIZON-EIC-2021-PATHFINDEROPEN-01-01 Grant No. 101046920 'iSenseDNA'), the Swedish Research Council (Grant No. 2021-05784, 2021-04629), and the Kempe Foundations (Grant No. JCK-3122). The authors thank M. Dipalo for the help with schematic Figure 1, E. Miele for the support in SERS data analysis, M. Ardini and G. Giovannini for the support on preliminary tests and fruitful discussions on photochemistry depositions, and the support from Nanofabrication Facility in IIT.

**Notes**
† The authors contributed equally to this work.
The authors declare no competing financial interest.

**Supporting information**
No supporting information is available for this manuscript.

# Plasmonic Photochemistry as a Tool to Prepare Metallic Nanopores with Controlled Diameter

*German Lanzavecchia[a,b†], Joel Kuttruff[c†], Andrea Doricchi[a,d], Ali Douaki[a], Kumaranchira Ramankutty Krishnadas[e], Isabel Garcia[e,f], Alba Viejo Rodríguez[g], Thomas Wågberg [h], Roman Krahne[a], Nicolò Maccaferri*[g,h], and Denis Garoli *[a]*

a. Optoelectronics Group, Istituto Italiano di Tecnologia, via Morego 30, I-16163, Genova, Italy.

b. Dipartimento di Fisica, Univesità di Genova, Via Dodecaneso 33, 16146 Genova,Italy

c. Universität Konstanz, Universitätsstr. 10, 78464 Konstanz, Germany

d. Dipartimento di Chimica, Univesità di Genova, Via Dodecaneso 31, 16146 Genova, Italy

e. CIC biomaGUNE, Basque Research and Technology Alliance (BRTA), 20014 Donostia-San Sebastián, Spain.

f. CIBER de Bioingeniería, Biomateriales y Nanomedicina, Instituto de Salud Carlos III

g. Department of Physics and Materials Science, University of Luxembourg, 106A Avenue de la Faïencerie, L-1511 Luxembourg, Luxembourg

h. Department of Physics and Umeå Centre for Microbial Research, Umeå University, SE-90187 Umeå, Sweden

*Corresponding authors: nicolo.maccaferri@umu.se; denis.garoli@iit.it;

**Note 1. Au nanopores – planar configuration - simulations**

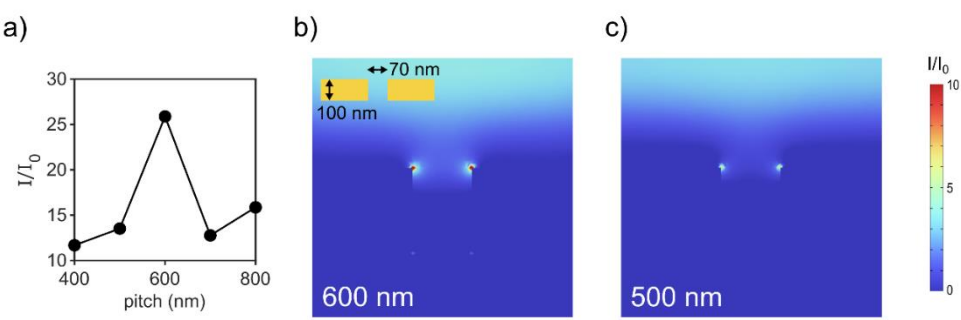

**Figure S1**. a) Intensity enhancement $I/I_0$ for a square gold nanohole array of varying pitch evaluated as average on the top surface of the hole using λ=638 nm. Hole diameter is 70 nm and film thickness is 100 nm. b),c) Intensity enhancement on (b) and off (c) fulfillment of the surface lattice resonance condition.

**Note 2. Au nanopores – planar configuration – Fabrication**

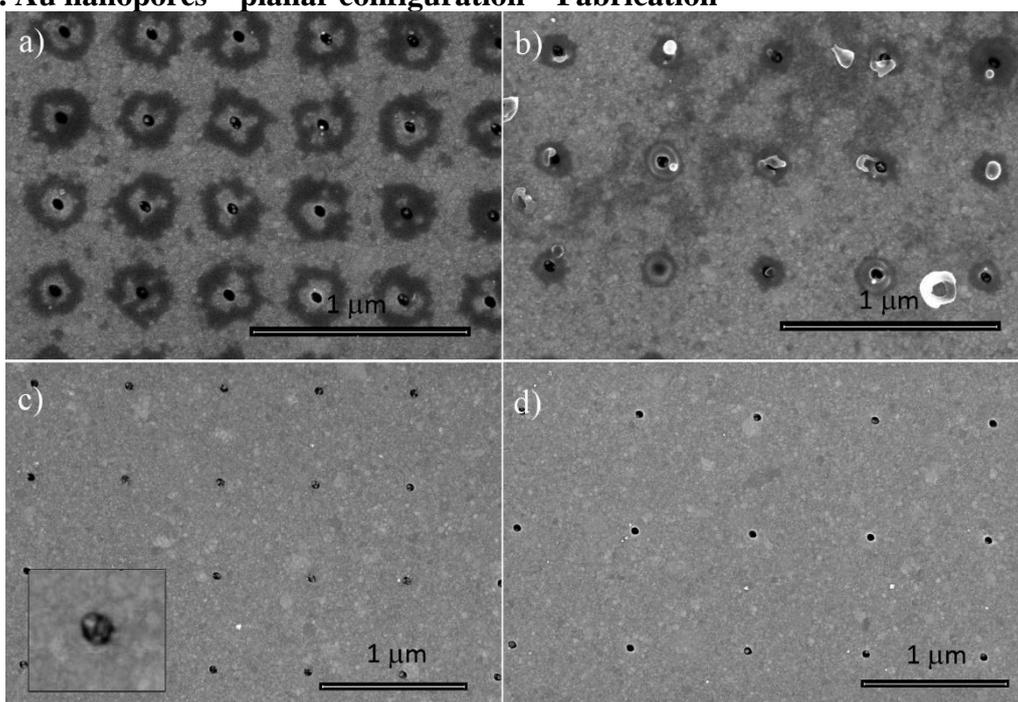

**Figure S2**. a) Nanopores array – pitch 400 nm. Ag seeding, 40 minutes under 638 nm laser excitation; b) Nanopores array – pitch 500 nm. Ag seeding, 40 minutes under 638 nm laser light excitation; c) Nanopores array – pitch 600 nm. Ag seeding, 40 minutes under 638 nm laser excitation; d) Nanopores array – pitch 800 nm. Ag seeding, 40 minutes under 638 nm laser light excitation.

**Note 3. Au-Ag nanorings on top of the pillars - fabrication**

Reduction of silver by sodium citrate in water:

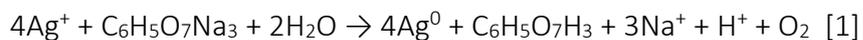

$4Ag^+ + C_6H_5O_7Na_3 + 2H_2O \rightarrow 4Ag^0 + C_6H_5O_7H_3 + 3Na^+ + H^+ + O_2$  [1]

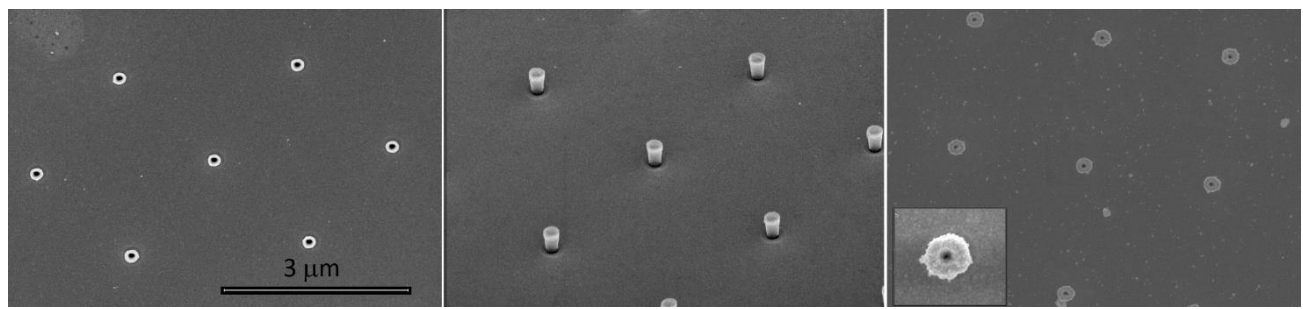

**Figure S3**. Examples of fabrication performed by using a ion current of 24pA (left panel bare Au ring – right panel after 40 minutes Ag seeding under white light).

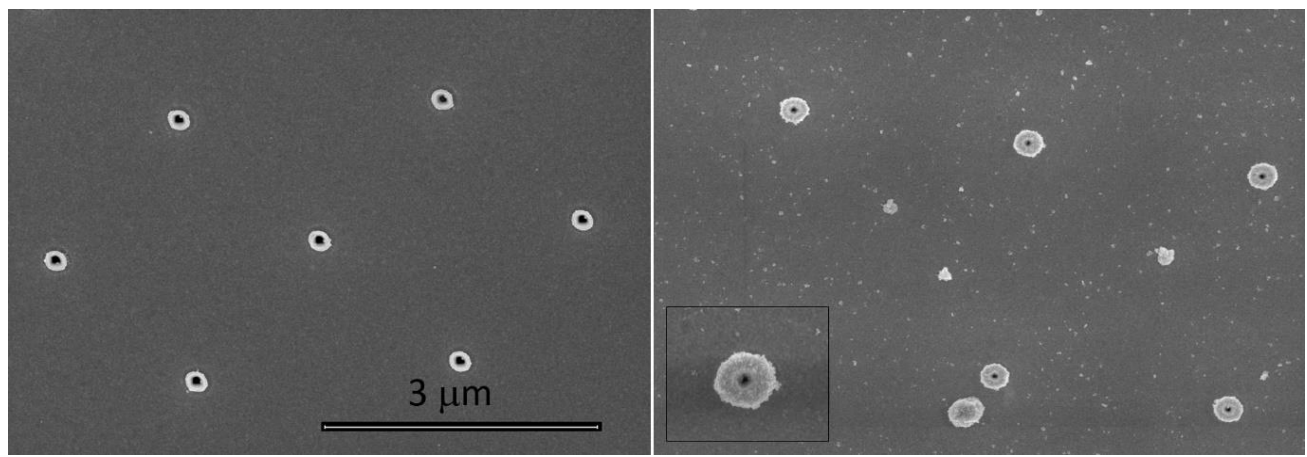

**Figure S4**. Examples of fabrication performed by using a ion current of 40pA (left panel bare Au ring – right panel after 40 minutes Ag seeding under white light).

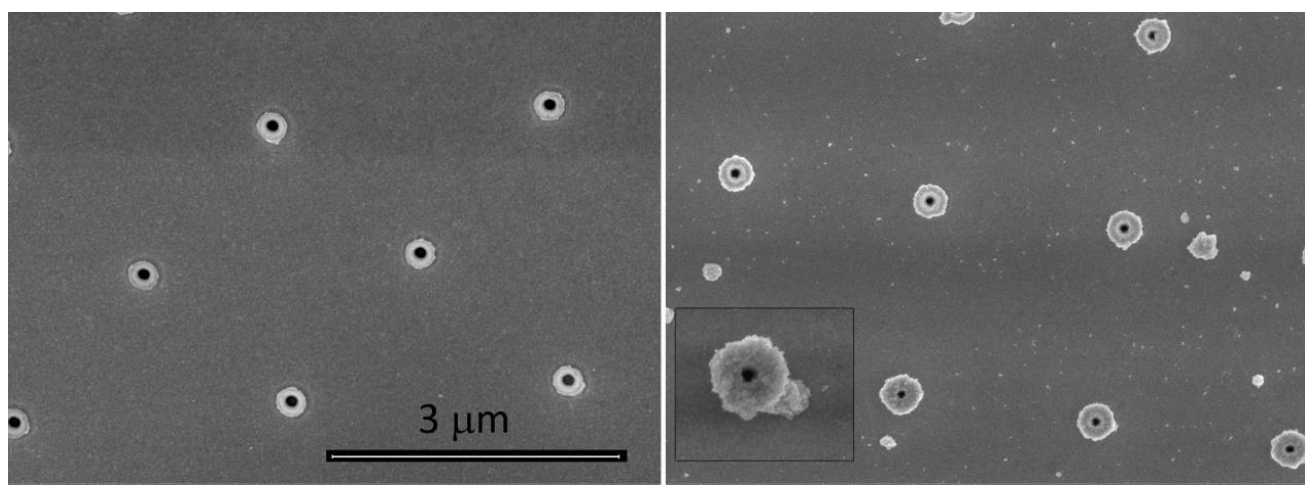

**Figure S5**. Examples of fabrication performed by using a ion current of 80pA (left panel bare Au ring – right panel after 40 minutes Ag seeding under white light).

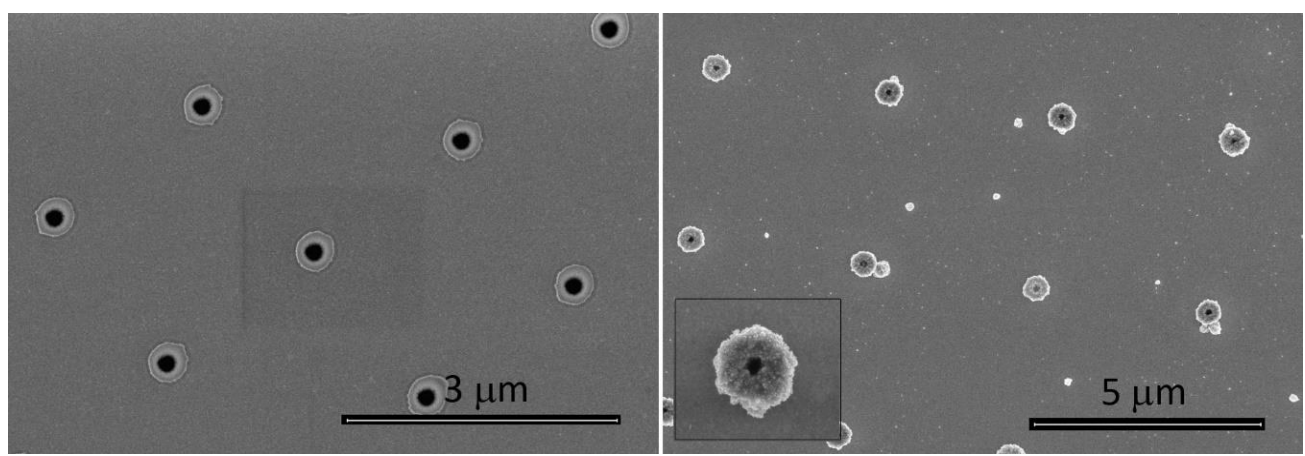

**Figure S6**. Examples of fabrication performed by using a ion current of 230pA (left panel bare Au ring – right panel after 40 minutes Ag seeding under white light).

## Note 4. Au ring on top of dielectric pillars under laser light.

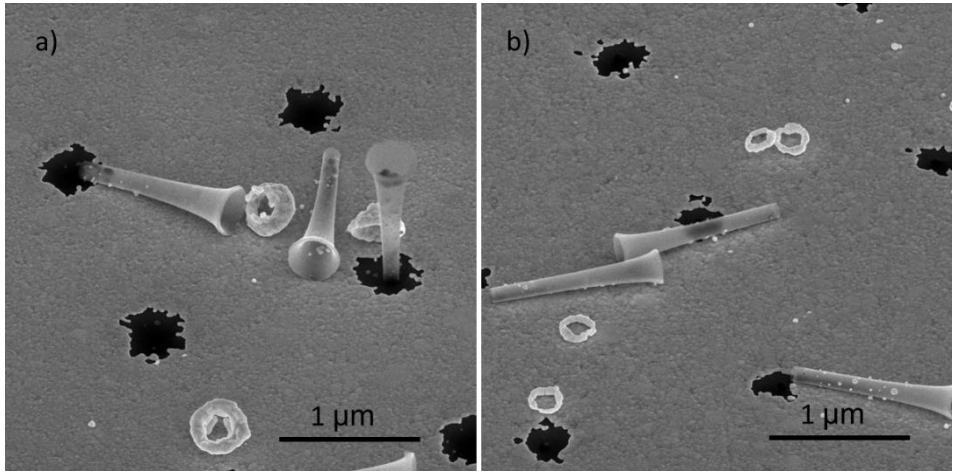

**Figure S7**. Samples destroyed during growth under laser illumination. a) Fabrication using 80 pA; b) Fabrication using 40 pA.

## Note 5. EDS spectrum of the Ag+Au nanoring

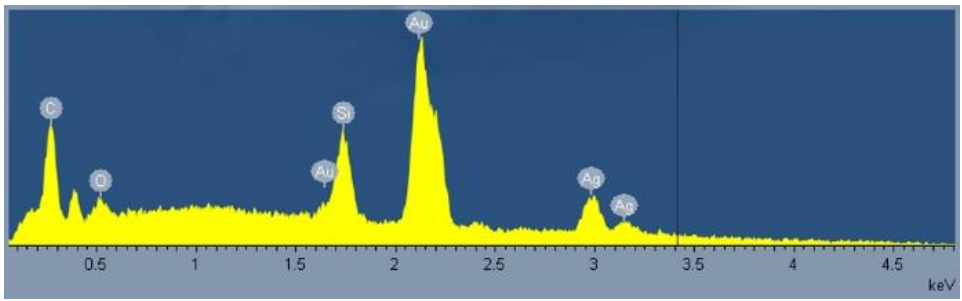

**Figure S8**. EDS spectrum of Ag-Au nanoring (corresponding to Fig. 2(d) main text).

## Note 6. Ag growth rate

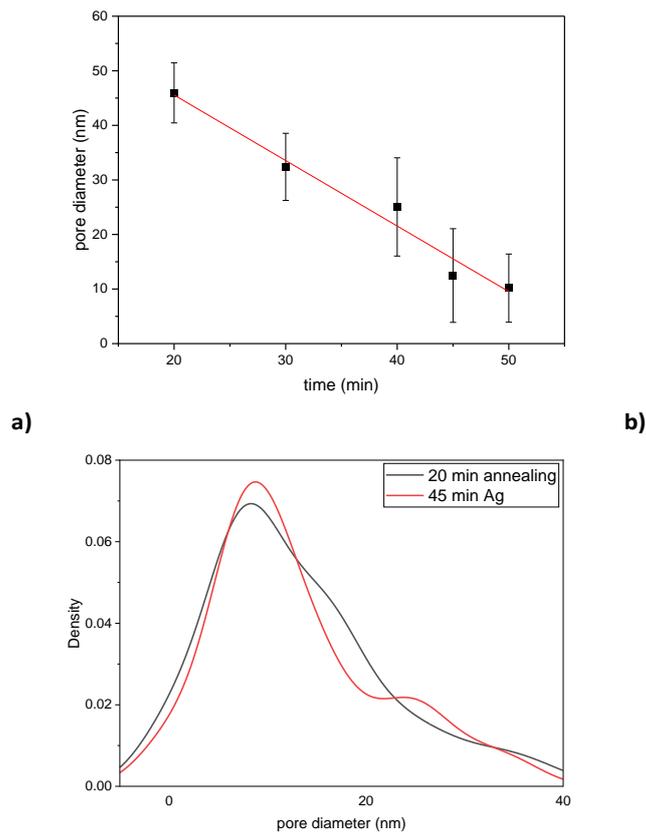

**Figure S9**. Plots of a) Pore diameter as a function of time and b) Pore diameter distribution density for an array of pores after 45 min deposition of Ag (red) and after 45 minutes deposition + annealing (black)

## Note 7. Fabrication of Au-Au nanoring – Au seeding with light – Self reduction.

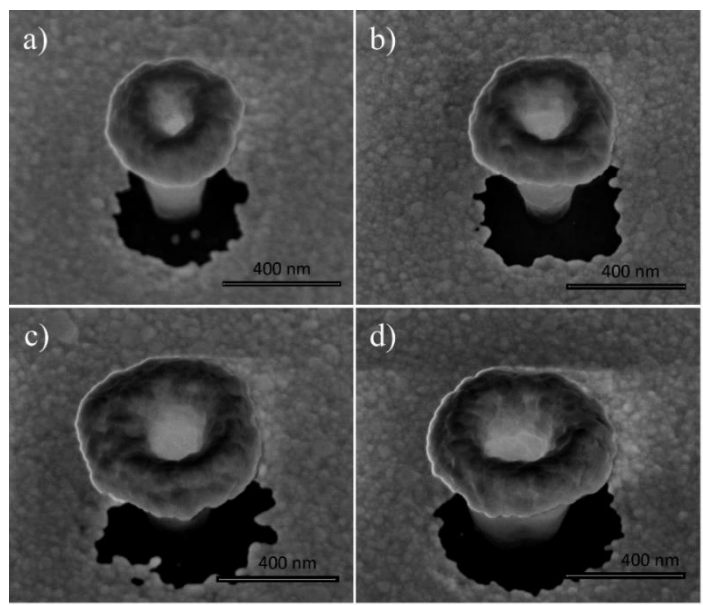

**Figure S10**. Fabrication of Au nanorings. Exposure for 40 minutes under white light excitation. a) Au ring prepared with a ion current of 24 pA with following seeding with Au salts and thermal annealing; b) Au ring prepared with a ion current of 40 pA with following seeding with Au salts and thermal annealing; c) Au ring prepared with a ion current of 80 pA with following seeding with Au salts and thermal annealing; d) Au ring prepared with a ion current of 230 pA with following seeding with Au salts and thermal annealing.

## Note 8. Fabrication of Au-Au nanoring – Au seeding with light - AuBr$_3$ + Citrate.

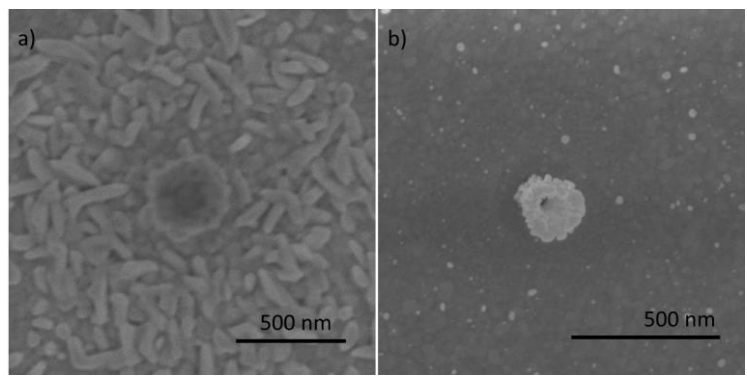

**Figure S11.** Au seeding on Au ring under white light for 15 min at a) 5mM and b) 0.05 mM

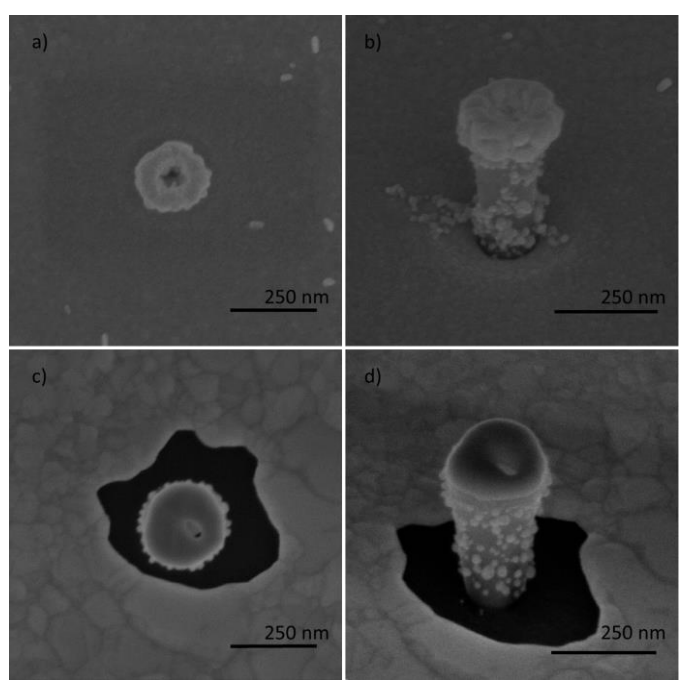

**Figure S12.** Au seeding on Au ring under white light for 15 min at 0.5 mM a-b) before and c-d) after 20 min annealing at 200°C.

## Note 9. Metal seeding on Au coated 3D pillars

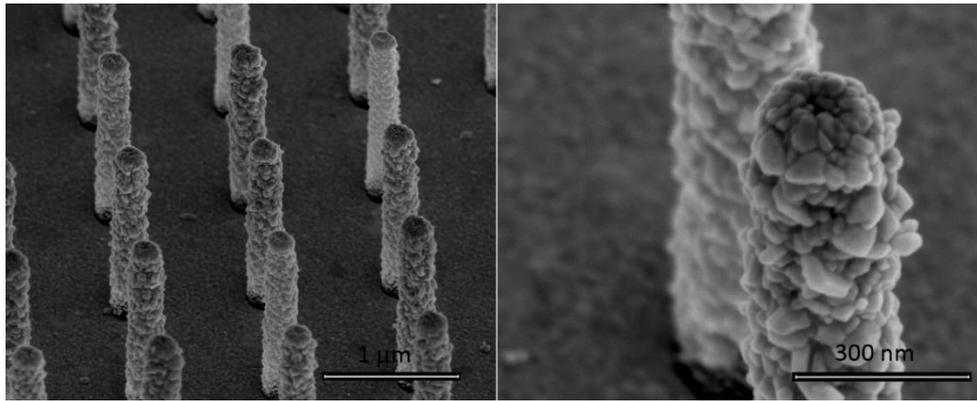

**Figure S13**. Plasmonic photochemistry – metal seeding with AgNO$_3$ under white light. Case of fully coated Au pillar.

**Note 10. Au ring on top of dielectric pillars – Ag growth with light stimuli – additional examples of growth and thermal annealing smoothing.**

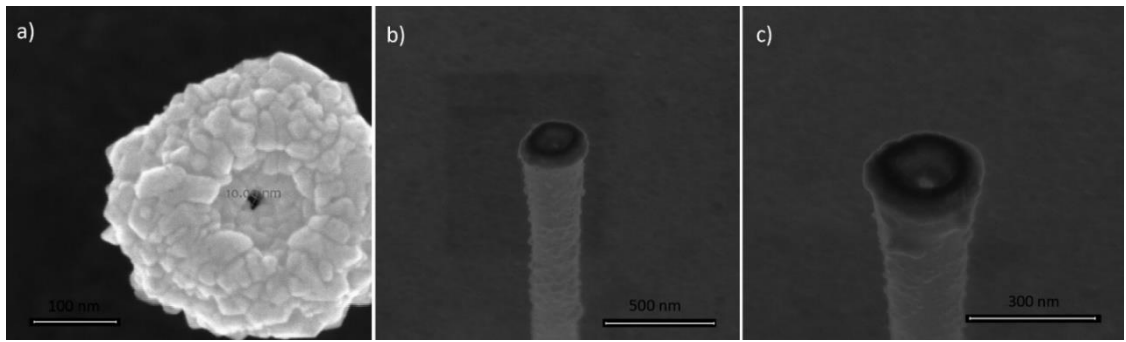

**Figure S14**. Ag seeding on Au ring. a) 40 minutes under white light; b) 10 minutes thermal annealing at 200 °C on a Ag ring prepared on an Au ring (24 pA); c) 10 minutes thermal annealing at 200 °C on a Ag ring prepared on an Au ring (40 pA);

**Note 11. Au ring on top of dielectric pillars – Ag growth without light stimuli**

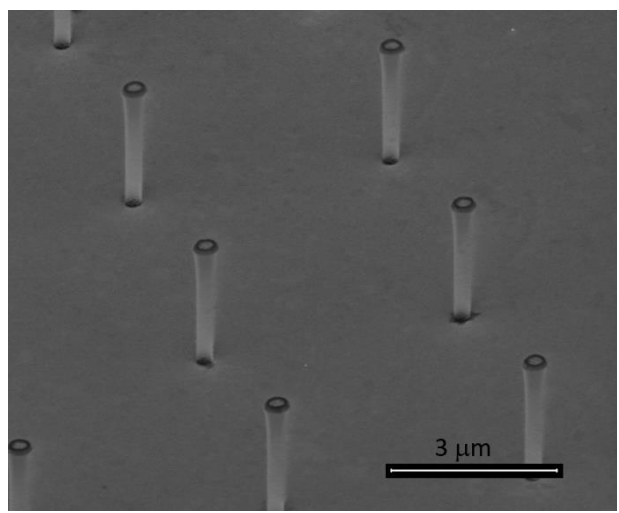

**Figure S15**. Ring Au antennae not exposed to light.

**Note 12. Au-Au nanorings**

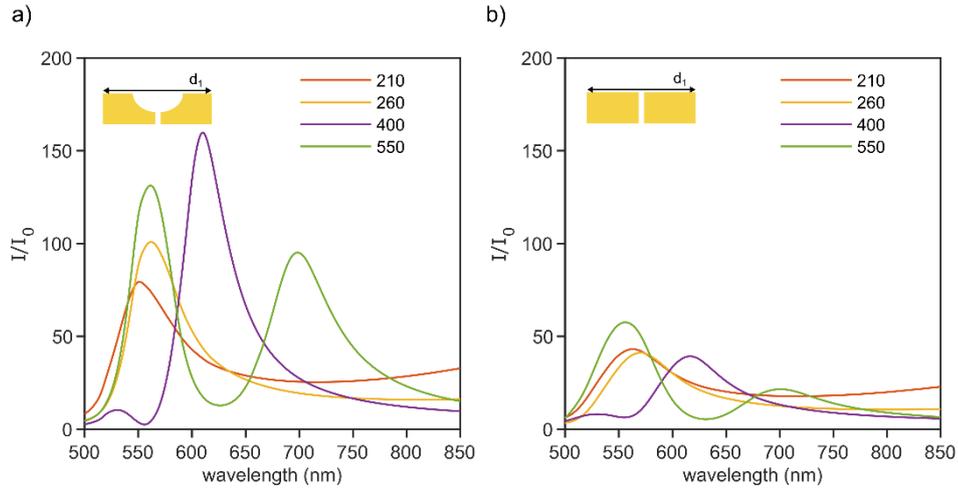

**Figure S16**. a) Electric field intensity enhancement $I/I_0$ for the curved geometry using different sizes $d_1/d_2$ as indicated in the inset. The enhancement factor is evaluated as average on the top surface of the nanopore. b) same as (a) for the planar geometry.

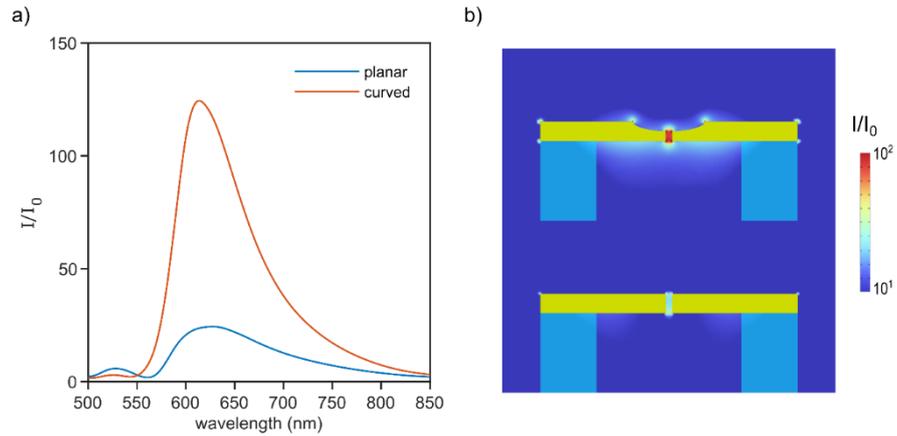

**Figure S17.** a) Electric field intensity enhancement $I/I_0$ for planar (blue) and curved (red) gold structures with outer diameter $d_1$=390 nm and inner diameter $d_2$=220 nm evaluated as average on the top surface of the nanopore. b) Field profiles at the resonances (curved structure, λ=610 nm, upper panel; planar structure, λ=630 nm, lower panel).

**Note 13. Pore diameter calculations.**

Pore diameter (d) was calculated according to the following equation:

$$G = \sigma \left[ \frac{4l}{\pi d^2} + \frac{1}{d} \right]^{-1}$$

using pore conductance (G), pore thickness (l), and solution conductivity (σ) [2]. Pore conductance (G) was calculated as G=I/V, using the mean current after stabilization of capacitive peaks, through a range of voltages [50 mV, -50 mV].

The pore thickness was set to about 100 nm for the starting 3D dielectric pillar-Au ring. For the final Ag nanopore the thickness was set in a range of expected metal thickness between 30 nm up to 150 nm with a final diameter between 4 and 8.9 nm.

The value of (σ) was set according to the literature for TE buffer.

**Note 14. DNA data at 200 kHz.**

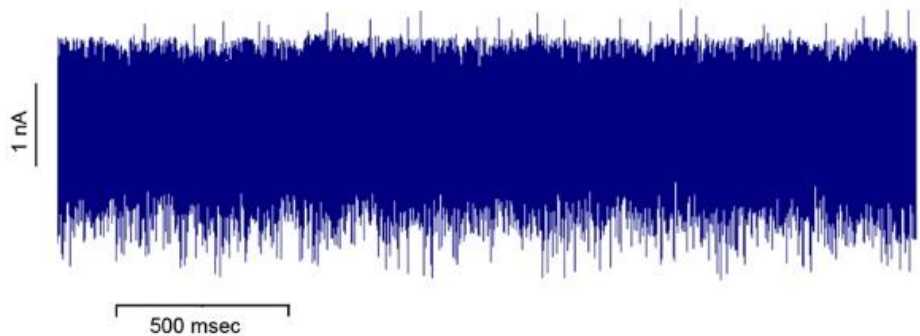

**Figure S18.** dsDNA translocations in 1 M KCl at V = 300 mV recorded at 200 kHz bandwidth using the portable 200 kHz nanopore reader from Elements SRL.

**Note 15. Au Nanoparticles.**

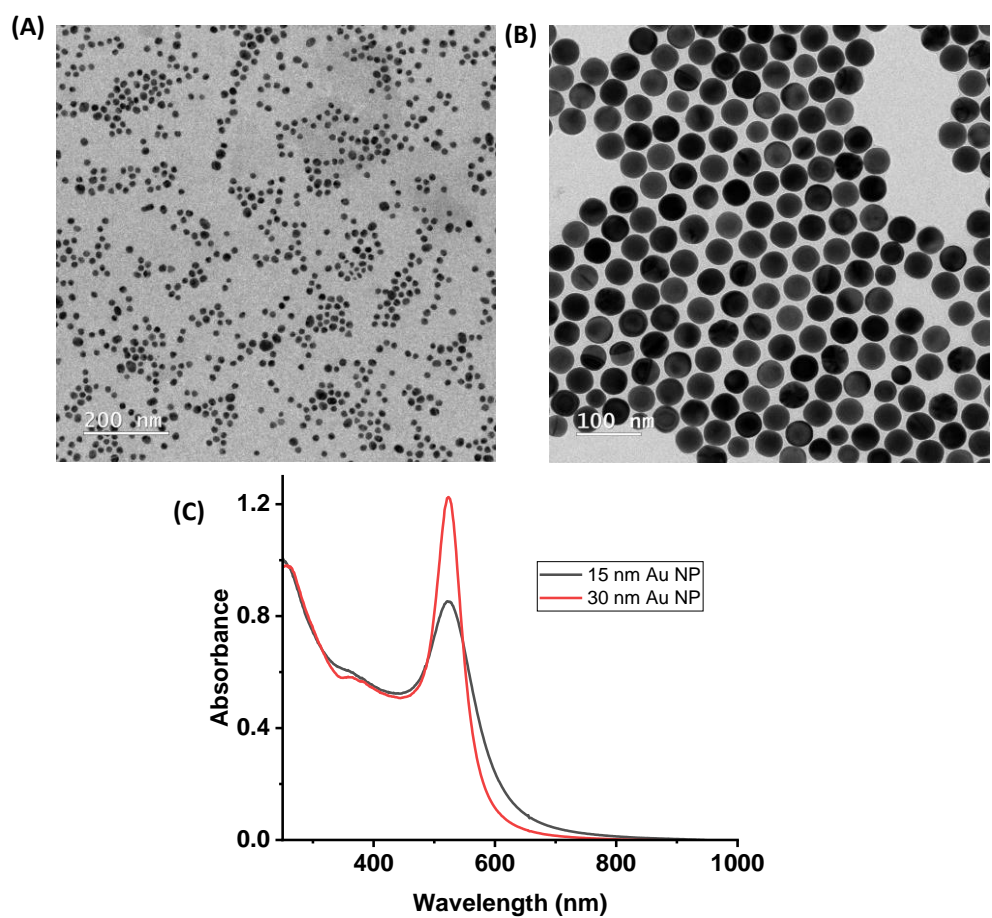

**Figure S19**. UV-Vis spectra and TEM images of PEGylated Au15 and Au30 NPs.

**Note 16. SERS analyses - Raman map on 3D nanopores array.**

Raman shift assignment used in the data analysis (Fig. 5(a) main text) [3]:

### dsDNA

| Main Peaks (cm$^{-1}$) | Assignments |
| --- | --- |
| 656; 652; 689 | 5-ring and 6-ring deformation |
| 776; 792; 798 | ring breath and 6-ring deformation |
| 1026; 1034 | rock $NH_3$; wag $CH_3$ |
| 1140; 1150 | bend N9-H, C8-H; stretch N1-C2; C2-N3 |
| 1220; 1229; 1240 | bend C8-H, N10-H11; ring stretch C-N |
| 1290; 1307 | ring stretch C-N, C-C |
| 1338; 1370; 1385 | ring stretch C-N, C-C; rock $NH_3$ |

   1461; 1490; 1482     ring stretch C-N; bend C8-H; stretch C4-N8

For R6G the 611 cm$^{-1}$ is the C–C–C ring in-plane bending vibration. 773 cm$^{-1}$ is the C–H out-of-plane bending vibration. 1127 cm$^{-1}$ is the C–H in-plane bending vibration. Peaks at 1310, 1360, 1507, 1573, and 1648 cm$^{-1}$ are assigned to the aromatic stretching vibrations.